\documentclass{article}

\usepackage{graphicx}
\usepackage{psfig}
\usepackage{epsfig}
\usepackage[round]{natbib}

\title{Use of the likelihood for measuring the skill of probabilistic forecasts}

\author{Stephen Jewson\footnote{\emph{Correspondence address}: RMS, 10 Eastcheap,
London, EC3M 1AJ, UK. Email: \texttt{x@stephenjewson.com}}\\
Risk Management Solutions, London, United Kingdom}

\begin{document}

\maketitle

\begin{abstract}

We define the likelihood and give a number of justifications for 
its use as a skill measure for probabilistic forecasts. 
We describe a number of different scores based on the likelihood, and
briefly investigate the relationships
between the likelihood, the mean square error and the ignorance.

\end{abstract}

\section{Introduction}
\label{introduction}

Users of forecasts need to know:

\begin{itemize}
  \item whether the forecasts they are receiving have been adequately calibrated
  \item whether the forecasts they are receiving are any better than an appropriate simple
  model such as climatology 
  \item which of the forecasts they are receiving is the best
\end{itemize}

To answer these questions, a single measure of forecast quality is needed. 
For calibration, the measure serves as a cost or benefit function that must be minimized or maximised in order
to find the optimum values for the free parameters in the calibration algorithm. 
For comparison with climatology or other forecasts, the measure serves as a way of deriving a ranking.

There are many standard measures of forecast quality.  
For example, for calibrating and comparing single-valued temperature forecasts, mean square error (MSE) is common.
For binary probabilistic forecasts, the Brier score~\citep{brier} is often used.
For continuous probability forecasts, the continuous rank probability score
and the ignorance have been suggested.

In this paper we 
will argue that likelihood-based measures provide a simple and natural
general framework for the evaluation of all kinds of probabilistic forecast.
For example, likelihood based measures can be used for binary and 
continuous probability forecasts, for temperature and precipitation, 
and for one lead time or many lead times simultaneously.

In section~\ref{probabilistic} we define the likelihood and discuss why we think it is a useful
measure of forecast skill.
In section~\ref{normal} we include
expressions for the likelihood for the normal distribution and in section~\ref{relations} we
discuss relations between the likelihood and other forecast scoring methods. Finally
in section~\ref{summary} we summarise and describe some areas of future work.

\section{Probabilistic forecasts and the likelihood}
\label{probabilistic}

How should we evaluate the skill of a probabilistic forecast?
We advocate the use
of a particular set of measures that are taken from classical statistics, and are all based on the 
\emph{likelihood}.
Likelihood is defined very simply as \emph{the probability of the observations given the forecast}.
In this phrase \emph{the observations} refers to the entire set of observations that we have available to validate a
certain forecast, and \emph{the forecast} refers to the entire set of corresponding forecasts.

Likelihood was first used by~\citet{fisher} as a method for fitting parameters to parametric distributions. 
Fisher proposed the likelihood as the natural benefit function that one should maximise in order to define the
best-fit parameters of the distribution. This suggestion was given a mathematical basis when it was
shown that the parameter values that maximise the likelihood are the most accurate possible estimates for the 
unknown parameters for most problems (see~\citet{casellab02}).

Fisher's problem, of how to evaluate the goodness of fit of a distribution to a number of samples,
is exactly the same as the problem of how to evaluate a probabilistic 
forecast. Instead of the distribution we have the probabilistic forecast
and instead of the samples we have the verifying observations. 

\subsection{Advantages of the likelihood as a measure for skill}

We consider that the likelihood has the following advantages as a measure of probabilistic forecast skill:

\begin{itemize}
  \item It has a simple definition that, from a purely intuitive point of view,
   seems to be a reasonable basis on which to compare forecasts
  \item It is mathematically optimal in the sense that estimates of parameters of calibration models fitted by maximising 
  the likelihood are usually the most accurate possible estimates (see~\citet{casellab02}).
  \item It is a generalisation to probabilistic forecasts of the most commonly used skill score for single forecasts: the RMSE 
  (see section~\ref{relations} below for a discussion of this). 
  \item It also shows how the RMSE can be generalised to the case of autocorrelated forecast errors
  \item The properties of the likelihood have been studied at great length over the last 90 years: it is well understood
  \item It is both a measure of resolution and reliability
  \item Likelihood can be used for both calibration and assessment: this creates consistency between these two operations
  \item Use of the likelihood also creates consistency with other statistical
  modelling activities, since most other statistical modelling uses the likelihood.
  This is important in cases where use of forecasts is simply a small part of a larger
  statistical modelling effort, as is the case for our particular business. 
  \item Likelihood can be used for all meteorological variables
  \item Likelihood can be used to compare multiple leads, multiple variables and multiple locations at the same time in a 
  sensible way (giving a single score) even when these leads, variables and locations are cross-correlated
\end{itemize}

\subsection{Forecast scores derived from the likelihood}

A number of different scores can be derived from the likelihood.

\begin{itemize}
  \item The log-likelihood (LL) reduces the range of values of the likelihood to a more manageable scale
  \item Minus the LL (MLL) has the characteristic that better forecasts have lower values: in this way it is analogous
  to the MSE
  \item The square root of the MLL (RMLL) has a further compressed scale
  \item All these measures can be transformed into skill scores from zero to one in the usual way
\end{itemize}

Other transformations are also possible: for instance, one might consider normalising by the number of data points.

\section{The likelihood for the normal distribution}
\label{normal}

For a normal distribution the likelihood is given by:

\begin{equation}\label{lik}
  L=\frac{1}{\sqrt{2 \pi \mbox{det}}}\mbox{exp}(-\frac{1}{2}(T-\mu)^T \Sigma^{-1} (T-\mu))
  \end{equation}

where $T$ is the vector of observations, $\mu$ is the vector of means from the forecast,
$\Sigma$ is the
covariance matrix of the forecast errors, and det is the determinant of $\Sigma$.

The log-likelihood is then:

\begin{equation}\label{loglik1}
  l=-\frac{1}{2} ln (2 \pi \mbox{det}) - \frac{1}{2}(T-\mu)^T \Sigma^{-1} (T-\mu)
\end{equation}

In the case where the forecast errors can be assumed to be uncorrelated in time, the likelihood becomes:

\begin{equation}\label{lik2}
  L=\frac{1}{\sqrt{2 \pi \mbox{det}}}\mbox{exp} 
    \left( - \frac{1}{2} \sum_{i=1}^{i=N} \frac{(T_i-\mu_i)^2}{\sigma_i^2} \right)
\end{equation}

and the log-likelihood is:
\begin{equation}\label{loglik2}
  l = -\frac{1}{2} \sum_{i=1}^{i=N} ln (2 \pi \sigma_i) 
    - \frac{1}{2} \sum_{i=1}^{i=N} \frac{(T_i-\mu_i)^2}{\sigma_i^2}
\end{equation}

When evaluating a forecast using the likelihood, calculating the covariance matrix is straightforward
because the forecast errors are known. When calibrating a forecast using the likelihood, calculating the
covariance matrix is more difficult. If it is reasonable to assume that the errors are uncorrelated
in time, then this simplifies the calibration considerably. However, this is generally not the case.

\section{Relations between the likelihood and other skill scores}
\label{relations}

Likelihood is closely related to the RMSE and the ignorance, as we see below.

\subsection{Relation between the likelihood and RMSE}

We show that the RMSE and the likelihood 
are consistent (i{.}e{.} give the same
ranking of forecasts) in the case of two normally distributed 
probabilistic forecasts with different means but 
the same constant spreads. Likelihood is used to compare the whole
distribution, while RMSE is used to compare the means.

Suppose we have two forecasts, $A$ and $B$, and suppose:
\begin{equation}
L_A > L_B 
\end{equation}
Taking logs, this gives:
\begin{equation}
l_A > l_B 
\end{equation}
Substituting in the expression for the log-likelihood for a normal distribution
we see that:
\begin{equation}
-\frac{N}{2}ln(2\pi)-\frac{N}{2}ln(\sigma)-\frac{1}{2 \sigma^2} \sum_{i=1}^{i=N} (x-f_A)^2 
> 
-\frac{N}{2}ln(2\pi)-\frac{N}{2}ln(\sigma)-\frac{1}{2 \sigma^2} \sum_{i=1}^{i=N} (x-f_B)^2 
\end{equation}

where $N$ is the number of observations, 
$f_a$ and $f_b$ are the time varying forecasts, and $x$ is the time-varying observations.

Cancelling terms from both
sides:
\begin{equation}
 -\frac{1}{2 \sigma^2} \sum_{i=1}^{i=N} (x-f_A)^2 > -\frac{1}{2 \sigma^2} \sum_{i=1}^{i=N} (x-f_B)^2 
\end{equation}

Cancelling more terms this gives:
\begin{equation}
 \sum_{i=1}^{i=N} (x-f_A)^2 < \sum_{i=1}^{i=N} (x-f_B)^2 
\end{equation}
or
\begin{equation}
  \mbox{MSE}_A < \mbox{MSE}_B
\end{equation}
and so we see that comparing these forecasts using likelihood or MSE gives the same results i{.}e{.} that
forecast A is better than forecast B.

\subsection{Relationship between the likelihood and ignorance}

\citet{roulstons02} describe a score for the assessment of probabilistic forecasts that they call
the \emph{ignorance}, and justify its usage on the basis of information theory and use in an
optimal betting strategy. They define the ignorance for a single forecast-observation pair as minus
the log (base 2) of the probability of the observation given the probabilistic forecast. 
We see that this is equivalent
to minus log (base 2) of the likelihood for that single forecast-observation pair.

Comparing forecasts using the ignorance or any of the likelihood-based scores described above will give
the same results if the forecasts errors are uncorrelated in time. If the errors are correlated in time,
and this is taken into account in the calculation of the likelihood, then they may give differing results.

One can consider the likelihood as a generalisation of the ignorance to a) forecasts with autocorrelated 
forecast errors and 
b) forecasts for many variables, locations or leads at once. One can consider the ignorance as a special
case of the likelihood when forecast errors are taken to be uncorrelated, and when looking at only a single
variable, location and lead.

\section{Summary}
\label{summary}

We have summarised the use of the likelihood for the evaluation of 
the skill of probabilistic forecasts. 
We believe that likelihood provides a useful general framework for the calibration and evaluation of
all probabilistic forecasts, for all variables.
We are in the process of applying the likelihood to various forecasting situations that are relevant to our
business: examples are given in~\citet{jewsonbz03a} and~\citet{jewsondh03a}.

A number of question arise that merit further investigation. These include:
\begin{itemize}
  \item When calibrating forecasts to maximise the likelihood, 
        what numerical methods can be used to estimate the forecast error covariance matrix?
  \item Is it really necessary to calculate the likelihood using the correct forecast error covariance matrix, or is it
  satisfactory in practice to make the assumption that forecast errors are uncorrelated? One can argue that 
  if the covariance matrix is not correctly modelled, then forecasts with autocorrelated errors are given 
  more credit than is their due. However, it may be that in practice the ranking of forecasts is the same whether
  or not the covariance is estimated accurately.
  \item What are the relationships, if any, between the likelihood and other skill scores apart from those
  discussed above?
\end{itemize}

\section{Acknowledgements}

The author has had helpful discussions on the topics discussed above with a number of people, including
Anders Brix, Pablo Doblas-Reyes, Renate Hagedorn and Christine Ziehmann.

\section{Version number}

This is version 2. Version 1 (August 12, 2003) contained some incorrect statements concerning the
relationship between the likelihood and the Brier score. Sorry for any confusion this may have caused.

\bibliography{jewson}

\end{document}